\documentclass[twoside,twocolumn,10pt]{revtex4-1} 

\usepackage[left=1in,right=1in,bottom=1in,top=1in]{geometry}
\usepackage{amsmath}
\usepackage{graphicx}  
\usepackage{amssymb}
\usepackage{units}
\usepackage{mathptmx}
\usepackage[T1]{fontenc}
\usepackage[latin9]{inputenc}
\usepackage{verbatim}
\usepackage{float}
\usepackage[colorlinks = true,
            linkcolor = blue,
            urlcolor  = blue,
            citecolor = blue,
            anchorcolor = blue]{hyperref}
\usepackage{epstopdf}

\begin{document}
\title{Power broadening effects on Electromagnetically Induced Transparency in $^{20}$Ne vapor}
\author{Boaz Lubotzky} 
\author{David Shwa}
\email[]{david.shwa@mail.huji.ac.il}
\author{Tao Kong}
\author{Nadav Katz}
\author{Guy Ron}
\affiliation{Racah Institute of Physics, The Hebrew University of Jerusalem, Jerusalem 91904, Israel }

\begin{abstract}We report here the first observation of electromagnetically induced transparency (EIT) in $^{20}$Ne. The power broadening of the EIT linewidth is measured as a function of neon pressure and RF excitation power. Doppler effects on the EIT broadening are found even at low pressures and low intensities, where the linewidth should be governed only by homogeneous effects.

\end{abstract}

\maketitle 

\section{Introduction}

Electromagnetically induced transparency (EIT) occurs when a three level system interacts with two resonant laser fields. This interaction results in a sharp transparency window for one of the beams in the otherwise absorptive media.  The spectral width of this window can be narrower than the atomic line width. EIT was first predicted in 1989 by Harris et al \cite{harris1997} and realized experimentally shortly after in Strontium vapor \cite{Bollerobservation} and in lead vapor \cite{Fieldobserve}. Due to its narrow spectral width EIT is a prominent candidate for applications in spectroscopy and magnetometry \cite{Lee98magnometer,Scully1992magnometer,Wyands99magnometer}. The narrow window of EIT is accompanied also by a steep dispersion leading to slow light \cite{Budker99Slowlight,Kasapi95Slowlight,NatureHau99slowlight,Schmidt96Slowlight,Khurgin:05SlowlightEIT}.    

EIT has been extensively investigated using alkali vapors such as rubidium \cite{Li_rubidium,MoseleyRubidium,olson:rubidium,XiaorubidiumEIT,Zhurubidium} and sodium \cite{Mitsunagasodium} using the relevant hyperfine levels. More recently EIT was also demonstrated on the Zeeman levels of metastable helium \cite{Ghoshhelium,Goldfarbhelium,KumarGoldfarbhelium}. Due to the zero nuclear spin of $^{4}He$, This system is unique for EIT research as it is a $\Lambda$ system with pure electronic spin, unlike alkali vapors where EIT is realized using hyperfine levels.        

Here we demonstrate for the first time EIT in metastable neon.  $^{20}Ne$ has some special features that separates it from helium or alkali vapors. Similarly to $^{4}He$, and unlike the alkali vapors, metastable neon is a closed two level system that possesses no nuclear spin. The metastability of the EIT ground state offers some unique advantages. First, it preserves the coherence of the medium due to de-excitation collisions. In this case the atom leaves the EIT system and lowers the total population without affecting the total coherence of the system. Another advantage is the possibility of controlling the population ratio of the metastable neon via changing the excitation mechanism. The stable neon atoms can potentially act as a buffer gas. Thus, no additional buffer gas is required.

Although most properties of metastable neon are shared also by helium, there are few distinctions. Neon has approximately five times larger atomic mass \cite{NISThandbook}, thus the EIT resonance width is less limited due to the transit time broadening. However, penning ionization in polarized metastable helium is suppressed by an additional two orders of magnitude relative to neon \cite{Vaasen2012penningionization}. Another difference is that the metastable neon system is suitable for Electromagnetically Induced Absorption (EIA) \cite{Lezama1999}, while metastable helium is a true EIT system. The reason for this difference lies in the angular momentum  of the two levels. In helium the transition is $J=1\rightarrow J'=0$ \cite{Goldfarbhelium}, thus the excited state has a lower angular momentum than the ground state. The transition in neon, on the other hand, is $J=2\rightarrow J'=3$. Degenerate two level systems exhibit EIA only when $J\rightarrow J'=J+1$ while EIT is possible using all the configurations \cite{Goren2004,Lipsich2000}.

\section{Experimental Setup}

\begin{figure}[t]
\centering
\centerline{\includegraphics[width=1\columnwidth]{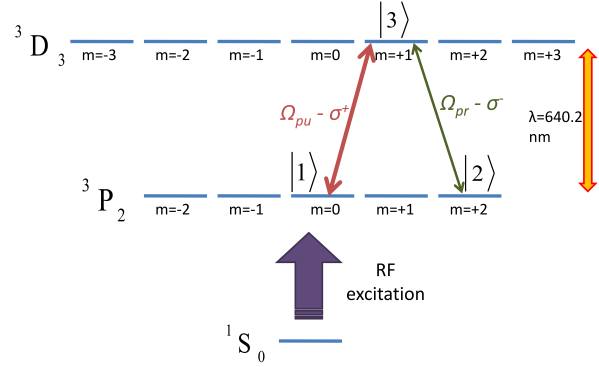}}
\caption{schematic energy level diagram for EIT in metastable neon.\cite{saloman:1113_neonwavelength,Ashmore2005}.
\label{fig:schematic-energy-level}}
\end{figure}

We use the neon $^{3}P_{2}\rightarrow$$^{3}D_{3}$ closed transition. This is a Zeeman degenerate two level system with energy splitting corresponding to a wavelength of 640.2 nm, where an EIT between two Zeeman sublevels can be employed. The ground metastable state $^{3}P_{2}$ has a total angular momentum quantum number of $J=2$ while the stable ground state $^{1}S_{0}$ has an angular momentum quantum number of $J=0$. Hence, an electric dipole transition is excluded and thus the $^{3}P_{2}$ level is a very long metastable state with approximately 15 seconds lifetime \cite{metastable_neonlifetime}. A schematic representation of the releveant neon energy levels is presented in Fig. \ref{fig:schematic-energy-level}.

\begin{figure}
\centering
\centerline{\includegraphics[width=1\columnwidth]{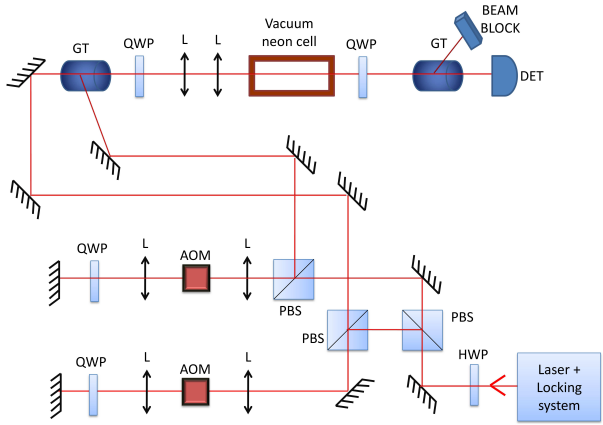}}
\caption{Experimental optical setup. HWP - half wave plate, QWP - quarter wave plate, L - lens, PBS - polarizing beam splitter, GT - Glan-Taylor polarizer, AOM - acousto optic modulator, DET - photodetector.  
\label{fig:setup} }
\end{figure}

Figure~\ref{fig:setup} shows our experimental setup for observing EIT in a hot metastable Ne vapor. The laser source is an amplified external cavity diode laser at $\sim$1280 nm which is converted to the transition wavelength by second harmonic generation. The laser is stabilized to the required transition using a standard saturation absorption locking setup \cite{Demtroeder2008}. The laser is split into pump and probe beams with different polarizations using a half wave plate. The frequency of each beam is controlled using an acousto-optic modulator in a double pass configuration to avoid frequency dependent beam deflections. After the frequency modulation the beams are combined using a Glan-Taylor calcite polarizer and pass through the neon cell spatially overlapping. Two quarter wave plates are inserted before and after the cell in order to change the polarization to circular relative to the beams propagation and back to linear polarization. For some measurements two lenses are inserted in order to control the beam diameter. After the neon cell the probe is separated from the pump using another Glan-Taylor polarizer and measured using a photodetector. The metastable neon excitation is achieved by a coaxial resonator with helical inner conductor \cite{coaxialresonators1959} that surrounds the cell and resonates at 61 MHz with an un-loaded Q factor of $\sim200$. We also use a pair of Helmholtz coils for axial magnetic field measurements. The vapor cell is shielded by a mu-metal housing reducing stray magnetic field by a factor of $\sim$100.

\section{Results}

\begin{figure}[t] 
\begin{centering}
\centerline{\includegraphics[width=1\columnwidth]{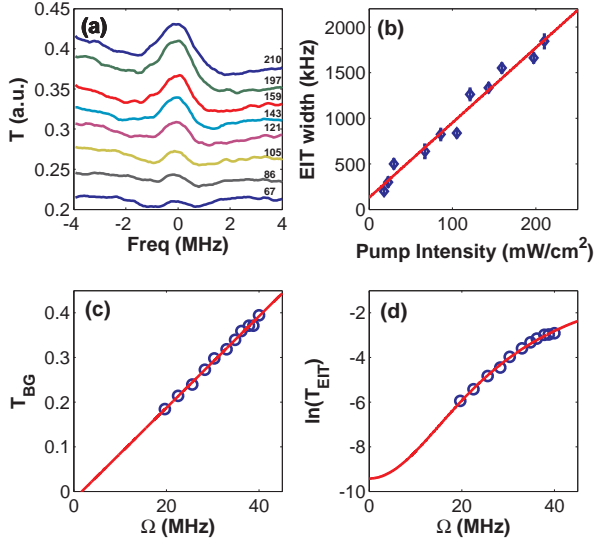}}
\par\end{centering}
\caption{(a) Probe transmission as a function of the two photon detuning for several pump intensities. The intensity on the right hand side is in mW/cm$^{2}$. (b) FWHM width of the EIT peak as a function of the pump intensity. The solid line is a linear fit to the data. (c) Background transmission ($-\ln{OD)}$) as a function of the Rabi frequency. Solid line is a linear fit to the data. (d) EIT Peak height as a function of the Rabi frequency. Solid line is a fit according to Eq. \ref{eq:peak_T}.}     
\label{fig:EIT}
\end{figure}

\begin{figure}[t] 
\begin{centering}
\centerline{\includegraphics[width=1\columnwidth]{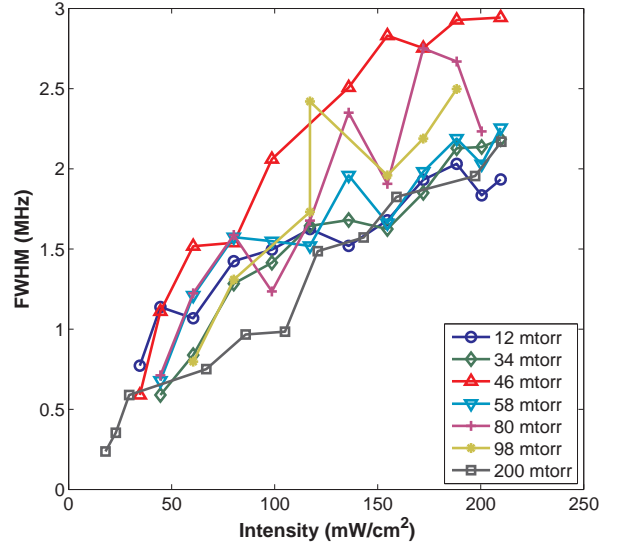}}
\par\end{centering}
\caption{Power broadening of the EIT for different neon pressures. The lines connecting the experimental data are guidance for the eye. The data for 200 mtorr is taken from Fig. \ref{fig:EIT}(b).}
\label{fig:pressure}
\end{figure}

Figure \ref{fig:EIT}(a) shows the probe transmission spectra for different pump intensities. Both pump and probe beams are approximately Gaussian with a waist of 1 mm. The optical power of the probe is 0.3 mW. Three major observations can be inferred from Fig. \ref{fig:EIT}(a). First, higher pump intensities result in an increase of the overall transmission due to hole burning. This is a broad feature due to the Doppler broadening. Near the two photon resonance we observe two opposite features. The first one is a broad (few MHz) absorption line. This is attributed to optical pumping of the atoms to the maximal Zeeman sublevel ($J=2,m_J=\pm2$). Due to the relevant Clebsch-Jordan coefficients this level has the highest probe absorption cross section $(g_f=1.66)$, thus an efficient pumping will cause an induced absorption. When the pump and the probe have the same wavelength the population is pumped more efficiently, thus an absorption profile with a homogeneous linewidth is created. The second feature is a narrow peak of enhanced transmission around the two photon resonance that is associated with EIT. Fig. \ref{fig:EIT}(b) shows the power broadening of the EIT peak. The theory of Doppler broadened EIT with no collisions predicts a quadratic function of the pump Rabi frequency $\Omega$ \cite{Ghoshhelium}: 
\begin{equation}
\Gamma_{EIT}=2\gamma_{g}+\frac{\Omega^{2}}{\delta}
\end{equation}

where $\gamma_{g}$ is the ground state decoherence and $\delta$ is an effective pumping rate that depends on the power of the pump. For low intensities such that $(\frac{\Omega}{W_{D}})^{2}\ll\frac{2\gamma_{g}}{\Gamma} $, where $W_{D}$ is the doppler width and $\Gamma$ is the natural decay rate, only a small resonant fraction of the atoms participate in the EIT. The effective rate in this case is $\delta_{low}=\Omega\sqrt{\nicefrac{\Gamma}{\gamma_{g}}}$. In the opposite regime, $(\frac{\Omega}{W_{D}})^{2}\gg\frac{2\gamma_{g}}{\Gamma} $, the intensity is strong enough to pump all the Doppler broadened population and the effective rate is $\delta_{high}=W_{D}$. In our experiment 10 MHz$<\Omega<$40 MHz, $\Gamma$=8 MHz and $W_{D}\simeq$1.3 GHz. The ground state decoherence is governed mostly by the transit-time broadening \cite{Thomas:84transittime}. Our beam has a width of $\sim$2 mm and the atomic ballistic velocity at room temperature is $\sim$500 m/s, thus the decoherence due to the transit-time can be estimated to be 250 kHz. Taking into account velocity changing collisions (VCC) with the stable neon atoms the atomic motion becomes diffusive, increasing the transit-time. For the pressure we use (200 mtorr) the diffusive transit-time is $\sim2.5$ longer than the ballistic one, making the decoherence $\sim$100 kHz. Using these values reveals that our system should be in the low intensity limit, hence the power broadening should be $\propto\sqrt{I}$. Strikingly, Fig. \ref{fig:EIT}(b) depicts a linear trend with the intensity. We fit the data to a linear function obtaining $\delta=$0.9 GHz which is very similar to $W_{D}$. It is important to note that Goldfarb et al. showed a similar effect in metastable helium that was attributed to many VCC that effectively explored all the Doppler velocity range \cite{Goldfarbhelium}. In their experiment they estimated $10^{4}$ collisions during the diffusion, but in our case due to smaller beam diameter and lower pressure the number of collisions is estimated to be $\sim5$. Figure \ref{fig:pressure} shows the EIT power broadening for several low pressure neon cells. It is apparent that even at very low pressures, where the number of collisions is $\ll1$ and the dynamics should be ballistic, the broadening is approximately linear, with $\delta\sim W_{D}$.

Another interesting aspect of EIT is its peak height. Figure \ref{fig:EIT}(d) shows the natural logarithm of the peak heights as a function of the Rabi frequency. The peak transmission can be estimated to be 

\begin{equation}\label{eq:peak_T} 
\ln{(\nicefrac{T}{OD})}=\frac{\ln{T_{0}}}{1+\frac{\Omega^{2}}{4W_{D}\gamma_{g}} },
\end{equation}

 where $OD=-\ln{T_{BG}}$ is the optical density of the medium with $T_{BG}$ taken from Fig. \ref{fig:EIT}(c) and $T_{0}$ is the transmission with no pump field.  This expression is similar to the one used in  \cite{Goldfarbhelium}, but with the exception of normalizing the transmission with the $OD$ due to the hole burning effect. The data was fitted to Eq. \ref{eq:peak_T} with the best fit presented as the solid line in Fig. \ref{fig:EIT}(d). The fit yields $4W_{D}\gamma_{g}=7\times10^{14}\,\mathrm{Hz}^{2} $ which is in good agreement with our estimation of the ground state decoherence.

\begin{figure}[t] 
\begin{centering}
\centerline{\includegraphics[width=1\columnwidth]{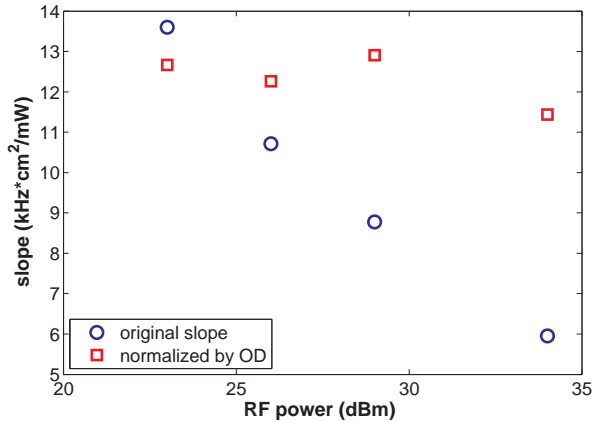}}
\par\end{centering}
\caption{Power broadening slope as a function of the RF power. blue circles - original slope. red squares - normalized slope by the optical density.}
\label{fig:rf}
\end{figure}

In our setup we can also control the RF power of the excitation, thus controlling the percentage of metastable neon. Figure \ref{fig:rf} shows a comparison of the power broadening slope for different RF powers (blue circles). This slope was calculated in the same way as in Fig. \ref{fig:EIT}(b). Higher RF powers have smaller slopes suggesting a narrowing. This narrowing can be explained when taking into consideration the effect of optical density such that \cite{Fleischauer2005} :

\begin{equation}\label{eq:ODfix}
\Gamma_{EIT}=2\gamma_{g}+\frac{\Omega^{2}}{\delta}\frac{1}{\sqrt{OD}}.
\end{equation} 
    
    We use Eq. \ref{eq:ODfix} for normalizing the slope by multiplying with $\sqrt{OD}$ (red squares). The results clearly show that the narrowing effect is due to the more effective population transfer to the metastable state at high RF powers.

\section{Discussion}
Although we describe here an EIT experiment, the full Zeeman level scheme of metastable neon as described in Fig. \ref{fig:schematic-energy-level} is not compatible with EIT but rather with EIA \cite{Lipsich2000,Goren2004}. This is true because the upper level has a higher angular momentum than the lower one, $J=2\rightarrow J'=3$. We did not observe EIA in our system but it is well known theoretically and experimentally that EIT appears in EIA systems under two conditions, a. when the degeneracy of the Zeeman sublevels is lifted. b. If the intensity of the pump is high enough such that $\Omega\gg\Gamma$ \cite{Goren2004,Kim2001,Lipsich2000}. All the results presented above were taken for high enough pump intensity such that the EIA should become EIT. On the other hand, when we lowered the intensity no EIA was discovered. Regarding the degeneracy, the main process that can lift it is the magnetic field due to the RF excitation. For the highest RF power we use, the magnetic field is estimated to be $\sim$1 G which is equivalent to $\sim1.5\,\mathrm{MHz}<\Gamma$. Hence the magnetic field is not enough to lift the degeneracy. Moreover, we also applied a constant axial magnetic field of up to $2\,G$ using an anti Helmholz configuration. This field shifted the two-photon EIT resonance linearly with the applied magnetic field as expected by the Zeeman splitting theory, but otherwise did not change any other EIT attributes. Possible resolution of this quandary (why EIA was not observed) may lie in the RF field. This field has a frequency which is much higher than $\Gamma$, thus the pump beam actually changes its magnetic axis causing a reduction in the effective coherence buildup between the EIA excited levels. This may cause a destruction of the EIA but not EIT, where only ground state coherence is needed.
 
Another unique result is the linear broadening with $\delta=W_{D}$ even at the low power limit and with low pressure such that VCC does not affect the dynamics. We did not find any simple explanation for this effect and it should be a part of further research in the future.

\section{Conclusion}

We present here the first observation of EIT in metastable neon. We measure linewidths as narrow as 200 kHz, significantly narrower than the natural linewidth. We measure the power broadening of the EIT linewidth for different pressures and RF excitation powers. The EIT width is inversely proportional to the Doppler width even for low pressures and intensities. By extrapolating the power broadening results we estimate the ground decoherence in our system to be 100 kHz. The dominant mechanism for this  decoherence is the transit time, which was also verified by the hard sphere model. Using the stable neon atoms as a buffer gas and increasing the VCC rate through higher pressures may lead to a narrower EIT that will be less restricted by the transit time broadening.

\section{Acknowledgments}

We wish to thank G. Arrad and A. Retzker for many fruitful discussions. We acknowledge support of Bikura grant 1567/12 and MAFAT. 

\bibliographystyle{apsrev4-1}
%

\end{document}